\documentclass[bibyear,rnote]{aa} 

\usepackage{graphicx}
\usepackage{txfonts}
\usepackage{hyperref}
\begin{document}
   \title{Overfitting and correlations in model fitting with separation ratios}
  \titlerunning{separation ratios}
   \author{Ian W. Roxburgh}
   \institute{Astronomy Unit, Queen Mary University of London, 
     Mile End Road, London E1 4NS, UK.
   \email {I.W.Roxburgh@qmul.ac.uk} }
   \date{Received  / Accepted  }
  \abstract{
  The $r_{01}$ and $r_{10}$ separation  ratios are not independent so combing them  into a single series $r_{010}$ is overfitting the data, this 
  can lead to almost singular covariance matrices with very large condition numbers, and hence to spurious results when comparing models and observations.
  Since the $r_{02}$ ratios are strongly correlated with $r_{10}$ and $r_{01}$ ratios, they should be combined into a single series  $r_{102}$ 
  (or $r_{012}$), which are not overfitted, and models and 
  observation compared using the covariance matrix $cov_{102}$ (or $cov_{012}$)  of the combined set.
   I illustrate these points by comparing the revised Legacy Project data with my results on the 10 Kepler stars in common.}

      \maketitle

\section{Introduction}

Frequency  separation ratios  are widely used in asteroseismic model fitting, i.e. finding models whose oscillation
properties match an observed set, as these ratios are almost independent of the outer layers of a star  (Roxburgh and Vorontsov 2003, 2013).

The ratios, constructed from frequencies $\nu_{n\ell}$ for angular degree  $\ell=0,1,2$, are customarily   defined as 
$$r_{01}(n)={(\nu_{n-1,0}-4\nu_{n-1,1}+6\nu_{n,0}-4\nu_{n,1}+\nu_{n+1,0})\over 8\, (\nu_{n,1}-\nu_{n-1,1})}~~~{\rm at}~~~\nu_{n.0}\eqno(1a)$$
$$r_{10}(n)=-{(\nu_{n-1,1}-4\nu_{n,0}+6\nu_{n,1}-4\nu_{n+1,0}+\nu_{n+1,1})\over 8\, (\nu_{n+1,0}-\nu_{n,0})}~~{\rm at}~\nu_{n,1}\eqno(1b)$$
   $$r_{02}(n) = {\nu_{n,0} - \nu_{n-1,2}\over \nu_{n,1}-\nu_{n-1,1}} ~~~~{\rm at}~~~~\nu_{n,0}\eqno(1c)$$

Since the ratios for given $n$ have several frequencies in common (eg $\nu_{n,0}, \nu_{n,1}$), as do ratios of neigbouring $n$, they are strongly correlated. 
When comparing model and observed values this requires one to match models and observed values of the ratios using the covariance matrices of the ratios of the observed values.  Care needs to be taken to ensure that one includes all the relevant correlations and that one does not overfit the data.

 \section{Combining $r_{01}$ and $r_{10}$ ratios into a single set $r_{010}$}
Several authors combine the observed ratios $r_{01}$ and $r_{10}$ into a single sequence $r_{010}$ (cf Silva Aguirre et al 2013, 2017), which should then be compared  with model values using the $r_{010}$  covariance matrix.  To demonstrate this can lead to anomalies I show in Fig 1 the fits of  the sequences $r_{01}, r_{10}, r_{010}$ for 16 Cyg A as given by the Legacy project in their revised MCMC analysis  (Lund  et al 2017a,b), to the very slightly different values (LegacyN) obtained directly from Eqns 1 using their frequencies. The fits for all 3 sequences using Legacy errors and for $r_{01}, r_{10}$  using the Legacy covariance matrices have $\chi^2\sim 10^{-3}$, whereas the fit for $r_{010}$ sequence has $\chi^2=4$. As shown in Table 1 similar results are obtained for the 10 stars in common between the Legacy Project and those analysed by myself (Roxburgh 2017). 

\newpage 

\begin{figure}[t]
\begin{center} 
    \includegraphics[width=9.cm]{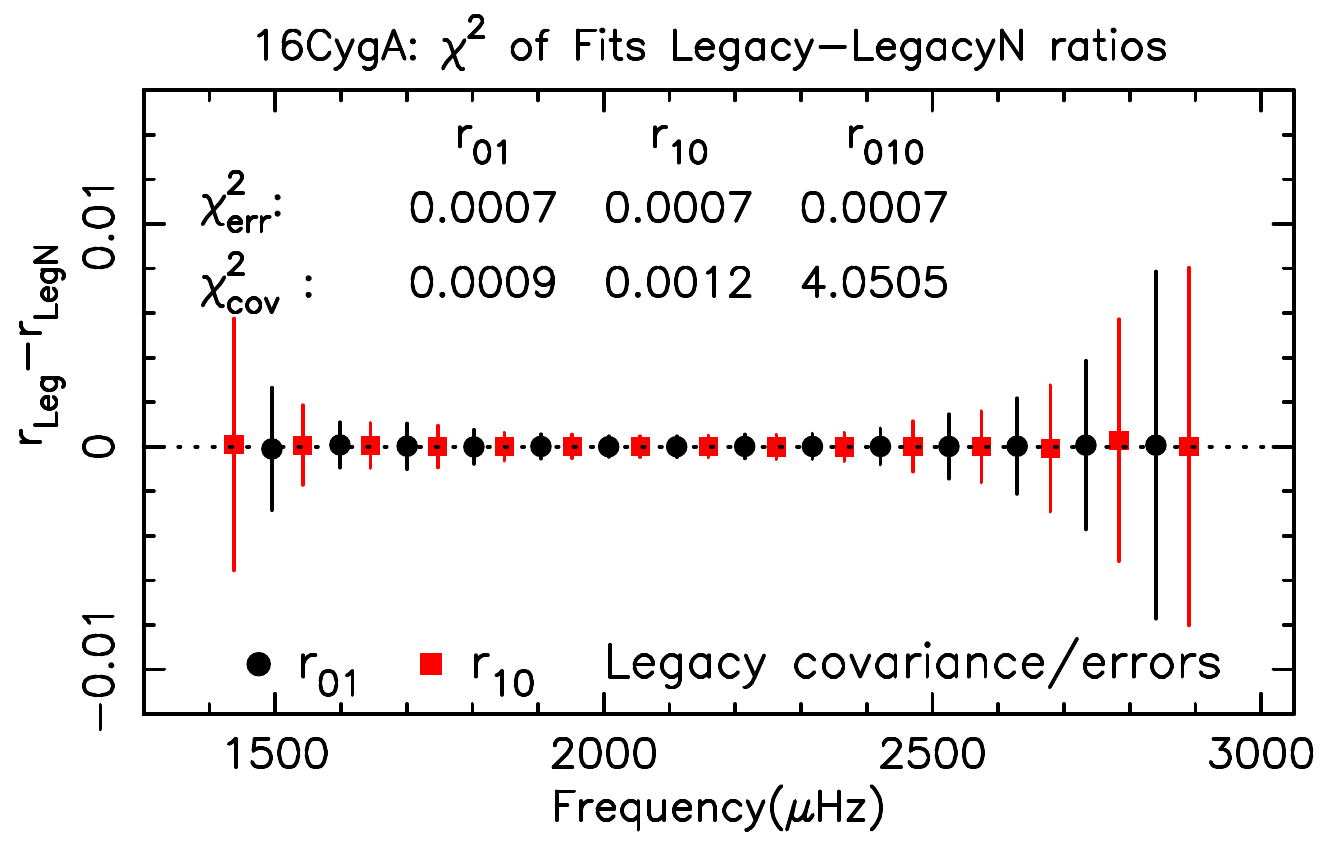}
  \vskip-5pt
   \caption{16CygA: Fits of Legacy project ratios to LegacyN values derived from the definitions and the Legacy freqenciest; $\chi^2_{err}$ using Legacy errors and 
 $\chi^2_{cov}$  using the Legacy covariance matrices.}

   \end{center}  
   \vskip-13pt
  \end{figure}

Table 2 gives the fits between Legacy ratios and my values of the ratios for these 10 stars, again using the revised Legacy  project covariance matrices (Lund 2017b).  Several have substantial differences between $\chi^2_{cov}$ for the $r_{010}$ sequence and
 $\chi^2_{cov}$  for the $r_{01}$ and $r_{10}$ sequences, while others show modest agreement. The result for 16 Cyg A is illustrated in Fig 2  

To identify the cause of this anomalous behaviour  I note that the frequencies can be be expressed as
(cf Roxburgh 2016)
$$\nu_{n,\ell} =\Delta [n+\ell/2 + \epsilon_\ell(\nu_{n,\ell})], ~~{\rm where}~~
\epsilon_\ell(\nu)=\alpha (\nu)-\delta_\ell(\nu)\eqno(2)$$
$\alpha(\nu)$ is the (almost) $\ell$ independent outer phase shift and the $\delta_\ell(\nu)$ are
the $\ell$ dependent inner phase shifts. The differences 
$$\epsilon_0(\nu)-\epsilon_\ell(\nu) =\delta_\ell(\nu)-\delta_0(\nu)\eqno(3)$$
at the same $\nu$, subtract out the surface phase shift giving a diagnostic of the stellar interior.
The ratios are interpolations in the $\epsilon_\ell(\nu_{n,\ell})$  which give approximations to the $\delta$ differences 
$$r_{01}\approx \delta_1 -\delta_0 ~~at~~ \nu_{n,0}, ~~~~~
r_{10}\approx \delta_1 -\delta_0 ~~at~~ \nu_{n,1}\eqno(4)$$
But these are not independent data.
\newpage

 \begin{table} [t]
\setlength{\tabcolsep}{11pt} 
\caption {Fits of  Legacy ratios and covariances to LegacyN  ratios }
\vskip -6pt
\small
\centering
 \begin{tabular}{l c c c c c c c c c c r c c c   } 
\hline\hline 
\noalign{\smallskip}
KIC no&$$&  $r_{01}$  &  $r_{10}$ &$r_{010}$ \\ [0.5ex]
\hline
\noalign{\smallskip}
3427720     &     $\chi^2_{cov}$     &     0.0006 &     0.0007 &     0.0424    \\[0.5ex]
6106415     &     $\chi^2_{cov}$     &     0.0002 &     0.0003 &     4.1139    \\[0.5ex]
6116048     &     $\chi^2_{cov}$     &     0.0006 &     0.0008 &     0.9903    \\[0.5ex]
6225718     &     $\chi^2_{cov}$     &     0.0011 &     0.0033 &     2.0442    \\[0.5ex]
6603624     &     $\chi^2_{cov}$     &     0.0004 &     0.0004 &     0.0524    \\[0.5ex]
8379927     &     $\chi^2_{cov}$     &     0.0011 &     0.0016 &     5.6099    \\[0.5ex]
8760414     &     $\chi^2_{cov}$     &     0.0009 &     0.0011 &     0.8123    \\[0.5ex]
9098294     &     $\chi^2_{cov}$     &     0.0007 &     0.0007 &     0.1008    \\[0.5ex]
10963065     &     $\chi^2_{cov}$     &     0.0004 &     0.0005 &     0.0266    \\[0.5ex]
12069424     &     $\chi^2_{cov}$     &     0.0009 &     0.0012 &     4.0505    \\[0.5ex]
12069449     &     $\chi^2_{cov}$     &     0.0014 &     0.0006 &     1.4373    \\[0.5ex]
\hline
\end{tabular}
\vskip-5pt
 \end{table}

 \begin{table} [h]
\setlength{\tabcolsep}{11pt} 
\caption {Fits of  Legacy ratios and covariances to Roxburgh's  ratios }
\vskip -6pt
\small
\centering
 \begin{tabular}{l c c c c c c c c c c r c c c   } 
\hline\hline 
\noalign{\smallskip}
KIC no&$$&  $r_{01}$  &  $r_{10}$ &$r_{010}$ \\ [0.5ex]
\hline
\noalign{\smallskip}
3427720     &     $\chi^2_{cov}$     &      0.345 &      0.411 &      0.545    \\[0.5ex]
6106415     &     $\chi^2_{cov}$     &      1.200 &      1.247 &      2.622    \\[0.5ex]
6116048     &     $\chi^2_{cov}$     &      0.697 &      0.648 &      1.048    \\[0.5ex]
6225718     &     $\chi^2_{cov}$     &      0.905 &      0.806 &      2.786    \\[0.5ex]
6603624     &     $\chi^2_{cov}$     &      0.217 &      0.130 &      0.302    \\[0.5ex]
8379927     &     $\chi^2_{cov}$     &      0.391 &      0.398 &     14.786    \\[0.5ex]
8760414     &     $\chi^2_{cov}$     &      0.751 &      0.660 &      3.836    \\[0.5ex]
9098294     &     $\chi^2_{cov}$     &      0.394 &      0.467 &      0.575    \\[0.5ex]
12069424     &     $\chi^2_{cov}$     &      1.043 &      1.057 &      8.813    \\[0.5ex]
12069449     &     $\chi^2_{cov}$     &      1.467 &      1.576 &     27.518    \\[0.5ex]
\hline
\end{tabular}
\vskip-11pt
 \end{table}

\begin{figure}[h]
\begin{center} 
    \includegraphics[width=9.cm]{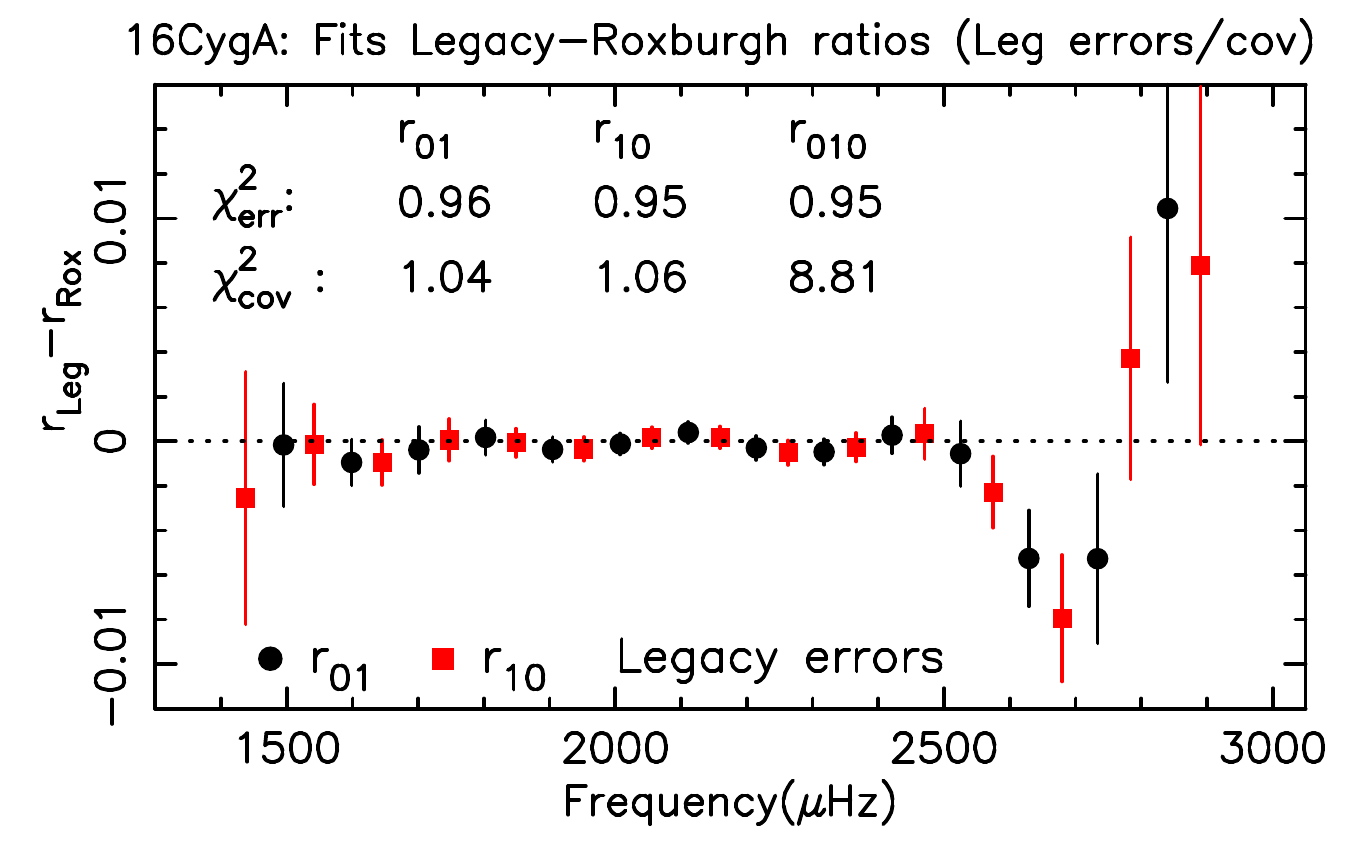}
  \vskip-5pt
   \caption{16CygA: Fits of Legacy project ratios to Roxburgh's values using Legacy errors and 
covariance matrices.  The covariance fit for $r_{010}$ is anomalously large.}
   \end{center}  
   \vskip-11pt
  \end{figure}

A simple illustration of this is to suppose one has $N$ independent values of $\epsilon_0(\nu)$ and $\epsilon_1(\nu)$  at the same frequencies $\nu_i$. Then since 
$\epsilon_0(\nu)=\alpha(\nu) -\delta_0(\nu)$ and  $\epsilon_1(\nu)=\alpha(\nu) -\delta_1(\nu)$, subtraction eliminates the $N$ values of $\alpha$ leaving $N$ independent values of $\delta_1-\delta_0$ at the $N$ frequencies $\nu_i$. One could determine additional values  of $\delta_1-\delta_0$ at any $\nu$, either by interpolating in 
the $N$ $\epsilon(\nu_i)$'s and subtracting, or equivalently  interpolating in the $N$ $\delta_1-\delta_0$ at $\nu_i$, but this does give any additional independent information.

The fact that with real data one needs to interpolate in  
either, or both the $\epsilon$'s  does alter this.  In principle  the values $\delta_1-\delta_0$ at $\nu_{n,1}$  can be derived by interpolation in the values at $\nu_{n,0}$, and likewise the values of $r_{10}$ by interpolation in the values of $r_{01}$. In this sense the combined sequence of $r_{010}$
which has $\sim 2N$ terms  is overdetermined.
\eject
Since such interpolation is dominantly linear it follows that the ratio  $r_{10} (n)$ is strongly correlated with the neighbouring values $r_{01} (n)$ and $r_{01} (n+1)$, and likewise for all neigbouring 
triplets. For Legacy 16 Cyg A, $corr\big\{ r_{10}(19), r_{01}(19)\big\} =0.84$ and $corr\big\{ r_{10}(19), r_{01}(20)\big\} =0.80$.  This in turn can lead to almost singular $r_{010}$ covariance matrices with large condition numbers (eg $\sim 1.6\,10^8$ for Legacy 16CygA, to be contrasted with values  883 and 587  the $r_{01}$, $r_{10}$ covariance matrices).

One could argue that the $r_{010}$ sequence is not overdetermined since one is simply comparing $2N$ frequencies; this is true, but one is comparing particular combinations of frequencies designed to eliminate the contributions of the outer layers and this introduces strong correlations between neighbouring terms which can lead to nearly singular covariance matrices.

As remarked by Lund et al (2017b) in their Erratum this leads to an inverse covariance matrix for $r_{010}$ with very large values oscillating in sign, which can lead to spurious values of the $\chi^2$ when comparing 2 sets of ratios. For example for 16 Cyg A  the elements on the  leading diagonal of the inverse covariance matrix are all positive with values up to $1.6\,10^{11}$ whereas the elements on the neigbouring diagonals have similar values but are all negative. 

 To illustrate that this behaviour is not just due to the properties of the Legacy covariance matrices I  show in Table 3 the comparisons for all 10 stars using my ratio covariance matrices. These show similar (but different) behaviour to those in Table 2.
\vskip3pt
Using the combined series  $r_{010}$ therefore can, (but may not) give spurious results.
One should avoid this possibility by comparing only one of the  $r_{01}$ or $r_{10}$
 sequences.  As discussed below this should be combined with the $r_{02}$ ratios.
 
  \begin{table} [t]
\setlength{\tabcolsep}{11pt} 
\caption {Fits of  Roxburgh's ratios and covariances to  Legacy  ratios }
\vskip -6pt
\small
\centering
 \begin{tabular}{l c c c c c c c c c c r c c c   } 
\hline\hline 
\noalign{\smallskip}
KIC no&$$&  $r_{01}$  &  $r_{10}$ &$r_{010}$ \\ [0.5ex]
\hline
\noalign{\smallskip}
3427720     &     $\chi^2_{cov}$     &      0.463 &      0.478 &      0.836    \\[0.5ex]
6106415     &     $\chi^2_{cov}$     &      1.219 &      1.278 &      7.442    \\[0.5ex]
6116048     &     $\chi^2_{cov}$     &      0.951 &      0.967 &      3.108    \\[0.5ex]
6225718     &     $\chi^2_{cov}$     &      0.781 &      0.660 &     11.064    \\[0.5ex]
6603624     &     $\chi^2_{cov}$     &      0.262 &      0.153 &      0.401    \\[0.5ex]
8379927     &     $\chi^2_{cov}$     &      0.489 &      0.471 &     16.825    \\[0.5ex]
8760414     &     $\chi^2_{cov}$     &      0.960 &      1.910 &      4.035    \\[0.5ex]
9098294     &     $\chi^2_{cov}$     &      0.433 &      0.557 &      1.230    \\[0.5ex]
10963065     &     $\chi^2_{cov}$     &      1.398 &      1.525 &      2.010    \\[0.5ex]
12069424     &     $\chi^2_{cov}$     &      1.041 &      2.539 &     11.968    \\[0.5ex]
12069449     &     $\chi^2_{cov}$     &      1.744 &      1.935 &      4.482    \\[0.5ex]
\hline
\end{tabular}
 \end{table}

\section{The combined sequence $r_{012}$ or $r_{102}$}
As is clear from the definitions of $r_{02}, r_{10}$ (or $r_{01}$) in Eqns 1 these ratios are strongly correlated as they have several frequencies in common,
one should therefore combine eg   $r_{10}, r_{02}$ into a combined sequence $r_{102}$ and compare observed and model values using the covariance matrix of
 the combined set. Or equally combine the ratios $r_{01}, r_{02}$ into a combined set $r_{012}$, but not both.
 Such combined sequences are not overfitted as from $3N$ values of  $\nu_{n,\ell}, \ell=0,1,2$, one can determine $\sim N$ independent values of each of the differences 
 $\delta_1 -\delta_0$, and $\delta_2 -\delta_0$.  The resulting covariance matrices have small condition numbers, eg for 16CygA $1.7\,10^3$ for $r_{102}$ as compared to $1.6\, 10^8$ for $r_{010}$. 
The $r_{102}$ fits for 16 Cyg A  are shown in Fig 3.
 
 \begin{figure}[t]
\begin{center} 
    \includegraphics[width=9.cm]{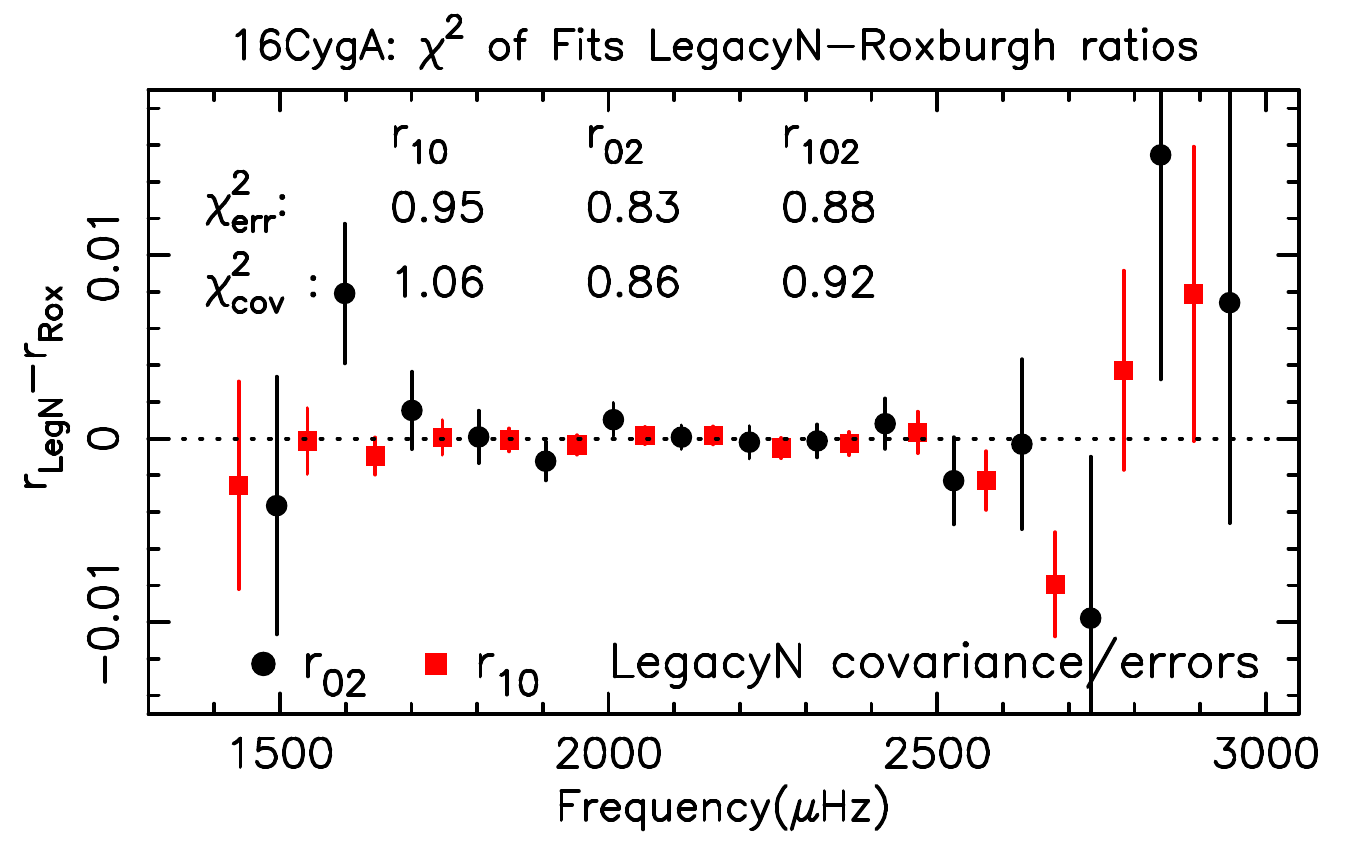}
  \vskip-5pt
   \caption{16CygA: Fits of Roxburgh's ratios to those of the Legacy project; $\chi^2_{err}$ using LegacyN errors and 
 $\chi^2_{cov}$  using the LegacyN covariance matrices.  The covariance and error fits for $r_{102}$ are of the same order and close to the average of the fits to  $r_{10},$ and $r_{02}$}
   \end{center} 
   \vskip-16pt 
  \end{figure}
  
Since the Legacy project does not give  $r_{102}$ or $r_{012}$  covariance matrices I generated these from the frequencies $\nu_k$ and frequency covariance matrices 
$cov(\nu_k,\nu_m)$ as given in Lund (2017b), the same procedure as was used to generate my covariance matrices.  The ratios and their derivatives with respect to the frequencies  follow directly from the definitions in Eqns 1 and
the covariance of any 2 ratios $r_i, r_j$  (be they $ r_{01}, r_{10}, r_{02}$) is given by
$$cov(r_i, r_j) = \sum_{ k}  \sum_{ m} {\partial r_i\over\partial \nu_k} cov(\nu_k,\nu_m)  {\partial r_j\over\partial \nu_m}\eqno(5)$$
This algorithm was used to determine the (LegacyN) covariance matrices of the  $r_{01}, r_{10}, r_{02}, r_{102}$ and $r_{012}$ sequences for all 10 stars in common to the Legacy project and my analysis,

The fits using the LegacyN covariance matrices are shown in Table 4. The fits for $r_{10}$ (and  $r_{01}$) are the almost the same as those given in Table 2 obtained using the covariance matrices supplied by the Legacy project from their MCMC analysis.  The values $\chi^2(r_{102})$  are consistent with the values  of $\chi^2(r_{10})$ and $\chi^2(r_{02})$, as are the values  for $\chi^2(r_{012})$  with the values  of $\chi^2(r_{01})$ and $\chi^2(r_{02})$

Table 5 shows the fits for the 10 stars using my ratio covariance matrices.  Again the values of $\chi^2(r_{102})$  are consistent with the values  of $\chi^2(r_{10})$ and $\chi^2(r_{02})$,
but are larger than those in Table 4, in particular the $\chi^2$ of the $r_{02}$ fits are considerably larger and consequently so too are the  $\chi^2(r_{102})$.
 This is a reflection of the frequency differences and considerably smaller errors on the  low frequencies from my MLE analysis as compared to those from the Legacy project's  MCMC analysis (see, for example, tables A1-A3 in Roxburgh 2017).

\section{Conclusions}
The ratios $r_{01}$, $r_{10}$ are not independent and in principle one set can be derived from the other by interpolation. From $N$ $\ell=0,1$ frequencies one 
can only derive $\sim N$ independent values of the  phase shift differences $\delta_1-\delta_0$ (which are approximated by the ratios). But the  $r_{010}$
 sequence has $\sim 2N$ components. In this sense one is overfitting the data.  Neighbouring elements of the covariance matrix can then be very
 strongly correlated leading to almost singular matrices with large condition numbers, and hence spurious results when comparing 2 sets of ratios.  
 Only one of  $r_{01}$, $r_{10}$ should be  used in model fitting.
 \newpage
Since the $r_{02}$, $r_{10}$ and $r_{01}$ ratios are correlated they should be combined into single sequence  $r_{012}$ or $r_{102}$ when comparing 2 sets of ratios. 
These sequences are not overfitted since
from  $N$ $\ell=0,1,2$ frequencies subtraction gives $\sim N$ values of both $\delta_1-\delta_0$ and $\delta_2-\delta_0$, which are approximated by the ratios $r_{10}, r_{02}$. The $r_{102}$ covariance matrices have reasonable condition numbers. 

 \begin{table} [t]
\setlength{\tabcolsep}{8pt} 
\caption {Fits of  LegacyN ratios and covariances to Roxburgh's  ratios }
\vskip -6pt
\small
\centering
 \begin{tabular}{l c c c c c c c c c c r c c c   } 
\hline\hline 
\noalign{\smallskip}
KIC no&$$&  $r_{10}$  &  $r_{02}$ &$r_{102}$ & $r_{012}$\\ [0.5ex]
\hline
\noalign{\smallskip}
3427720     &     $\chi^2_{cov}$     &      0.427 &      0.772 &      0.582 &      0.580    \\[0.5ex]
6106415     &     $\chi^2_{cov}$     &      1.198 &      2.507 &      2.206 &      2.188    \\[0.5ex]
6116048     &     $\chi^2_{cov}$     &      0.657 &      0.597 &      0.562 &      0.567    \\[0.5ex]
6225718     &     $\chi^2_{cov}$     &      0.830 &      1.397 &      1.060 &      1.042    \\[0.5ex]
6603624     &     $\chi^2_{cov}$     &      0.138 &      1.926 &      1.162 &      1.226    \\[0.5ex]
8379927     &     $\chi^2_{cov}$     &      0.368 &      2.958 &      1.576 &      1.544    \\[0.5ex]
8760414     &     $\chi^2_{cov}$     &      0.661 &      2.831 &      1.742 &      1.550    \\[0.5ex]
9098294     &     $\chi^2_{cov}$     &      0.481 &      0.244 &      0.430 &      0.402    \\[0.5ex]
12069424     &     $\chi^2_{cov}$     &      1.059 &      0.857 &      0.919 &      0.947    \\[0.5ex]
12069449     &     $\chi^2_{cov}$     &      1.577 &      1.022 &      1.360 &      1.225    \\[0.5ex]
\hline
\end{tabular}
 \end{table}
 
  \begin{table} [t]
\setlength{\tabcolsep}{8pt} 
\caption {Fits of  Roxburgh's ratios and covariances to  LegacyN  ratios }
\vskip -6pt
\small
\centering
  \begin{tabular}{l c c c c c c c c c c r c c c   } 
\hline\hline 
\noalign{\smallskip}
KIC no&$$&  $r_{10}$  &  $r_{02}$ &$r_{102}$ & $r_{012}$\\ [0.5ex]
\hline
\noalign{\smallskip}
3427720     &     $\chi^2_{cov}$     &      0.489 &      0.683 &      0.544 &      0.536    \\[0.5ex]
6106415     &     $\chi^2_{cov}$     &      1.270 &      2.762 &      2.259 &      2.361    \\[0.5ex]
6116048     &     $\chi^2_{cov}$     &      0.950 &      0.735 &      0.783 &      0.793    \\[0.5ex]
6225718     &     $\chi^2_{cov}$     &      0.663 &      1.045 &      0.859 &      0.924    \\[0.5ex]
6603624     &     $\chi^2_{cov}$     &      0.162 &      4.139 &      2.591 &      2.798    \\[0.5ex]
8379927     &     $\chi^2_{cov}$     &      0.441 &      1.069 &      0.701 &      0.646    \\[0.5ex]
8760414     &     $\chi^2_{cov}$     &      1.938 &     10.322 &      5.557 &      5.523    \\[0.5ex]
9098294     &     $\chi^2_{cov}$     &      0.564 &      0.268 &      0.479 &      0.425    \\[0.5ex]
10963065     &     $\chi^2_{cov}$     &      1.524 &      1.821 &      1.535 &      1.496    \\[0.5ex]
12069424     &     $\chi^2_{cov}$     &      2.607 &      6.608 &      4.697 &      4.790    \\[0.5ex]
12069449     &     $\chi^2_{cov}$     &      1.935 &      6.090 &      4.883 &      3.965    \\[0.5ex]
\hline
\end{tabular}
 \end{table}

\end{document}